\documentclass[fleqn,twoside]{article}
\usepackage{espcrc2}
\usepackage{latexsym}
\usepackage[dvips]{color}

\input epsf.sty
\newdimen\psfigsize

\def\etal{{\it et al.}}

\def\BE{\begin{equation}}
\def\EE{\end{equation}}
\def\BEA{\begin{eqnarray}}
\def\BEAN{\begin{eqnarray*}}
\def\EEA{\end{eqnarray}}
\def\EEAN{\end{eqnarray*}}

\definecolor{Red}           {cmyk}{0,1,1,0}
\definecolor{Blue}          {cmyk}{1,1,0,0}
\definecolor{Green}         {cmyk}{1,0,1,0}

\title{Light hadron properties with improved staggered quarks}

\author{
    C. Bernard \address{Department of Physics, Washington University, St.~Louis, MO 63130, USA},
    T. Burch \address[ARIZ]{Department of Physics, University of Arizona, Tucson, AZ 85721, USA},
    T. DeGrand \address{Physics Department, University of Colorado, Boulder, CO 80309, USA},
    C. DeTar \address[UTAH]{Physics Department, University of Utah, Salt Lake City, UT 84112, USA},
    Steven Gottlieb \address{Department of Physics, Indiana University, Bloomington, IN 47405, USA},
    E.B. Gregory \addressmark[ARIZ],
    U.M. Heller \address{CSIT, Florida State University, Tallahassee, FL 32306-4120, USA},
    J. Osborn \addressmark[UTAH],
    R. Sugar \address{Department of Physics, University of California, Santa Barbara, CA 93106, USA},
    and
    D.~Toussaint \addressmark[ARIZ]\thanks{Presented by D. Toussaint}
}

\begin{document}
\begin{abstract}
Preliminary results from simulations with 2+1 dynamical quark flavors
at a lattice spacing of 0.09 fm are combined with earlier results at
$a=0.13$ fm.  We examine the approach to the continuum limit and
investigate the dependence of the pseudoscalar masses and decay
constants as the sea and valence quark masses are
separately varied.
\end{abstract}
\maketitle

\renewcommand{\thefootnote}{\fnsymbol{footnote}}

\begin{figure}[t]
\rule{0.0in}{0.3in}\vspace*{-0.3in}\\
\epsfxsize=3.00in
\epsfbox[0 0 4096 4096]{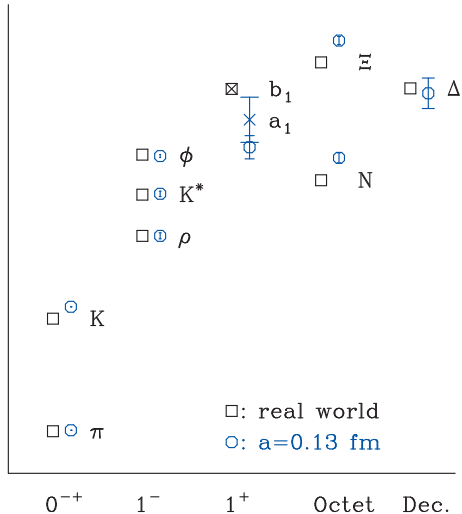} 
\rule{0.0in}{0.01in}\vspace{-0.5in}\\
\caption{
2+1 flavor lattice spectrum results compared with the real world.
These are from simulations with $a\approx 0.13$ fm (no continuum
extrapolation), with a simple linear extrapolation to the physical
quark mass.
\label{BIGPIC_FIG}
}
\end{figure}

\begin{figure}[t]
\rule{0.0in}{0.3in}\vspace*{-0.3in}\\
\epsfxsize=3.00in
\epsfbox[0 0 4096 4096]{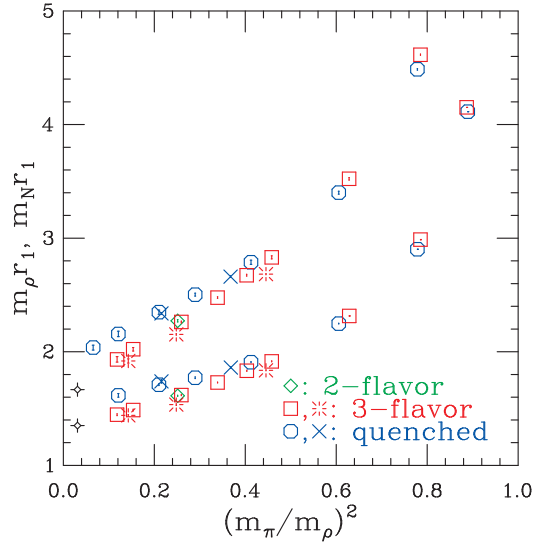} 
\rule{0.0in}{0.01in}\vspace{-0.5in}\\
\caption{
Nucleon and $\rho$ masses in units of $r_1$ (upper
and lower bands, respectively).
The open symbols are $a=0.13$ fm.
Bursts and crosses are $a=0.09$ fm (preliminary).
Fancy plusses are physical values, using $r_1=0.35$ fm.
\label{MRHONUC_FIG}
}
\end{figure}


\begin{figure}[t]
\rule{0.0in}{0.3in}\vspace*{-0.3in}\\
\epsfxsize=3.00in
\epsfbox[0 0 4096 4096]{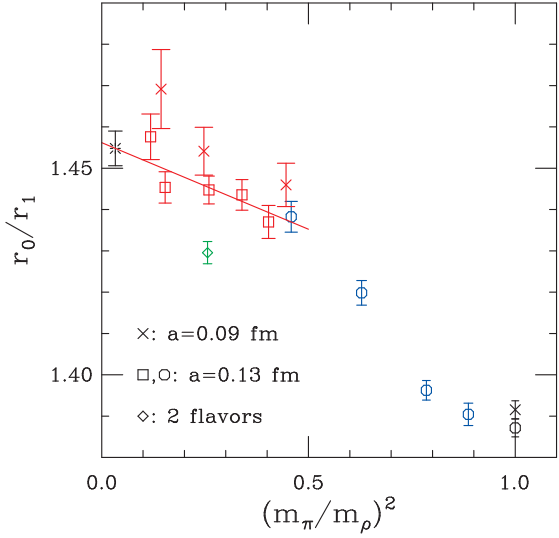} 
\rule{0.0in}{0.01in}\vspace{-0.5in}\\
\caption{
A shape parameter for the static quark potential.
Octagons and squares are 0.13 fm runs for
three equal masses and 2+1 flavors respectively.
The crosses are preliminary 0.09 fm results.
The burst is an extrapolation to the physical value.
of $(m_\pi/m_\rho)^2$.
\vspace{-0.2in}
\label{POTSHAPE_FIG}
}
\end{figure}




\begin{figure}[t]
\rule{0.0in}{0.3in}\vspace*{-0.3in}\\
\epsfxsize=3.00in
\epsfbox[0 0 4096 4096]{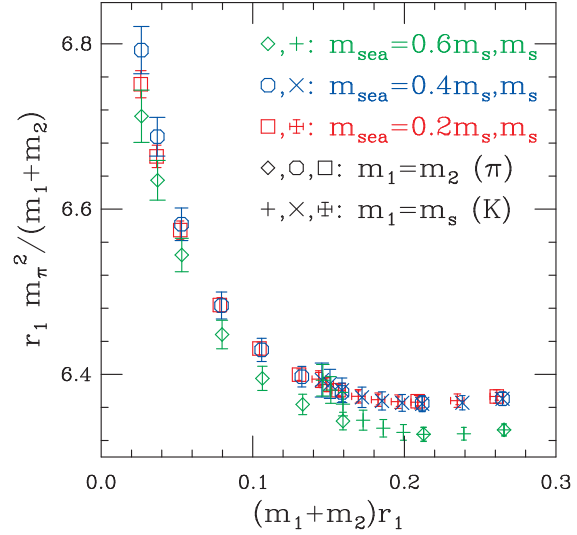} 
\rule{0.0in}{0.01in}\vspace{-0.5in}\\
\caption{
\label{MPISQ_PQ_FIG}
Partially quenched $M_{PS}^2/(m_1+m_2)$ for $a \approx 0.13$ fm.
$m_1$ and $m_2$ are valence quark masses.
Light sea quark masses are 0.2, 0.4 and 0.6 $m_s$.
\vspace{-0.1in}
}
\end{figure}

\begin{figure}[t]
\rule{0.0in}{0.3in}\vspace*{-0.3in}\\
\epsfxsize=3.00in
\epsfbox[0 0 4096 4096]{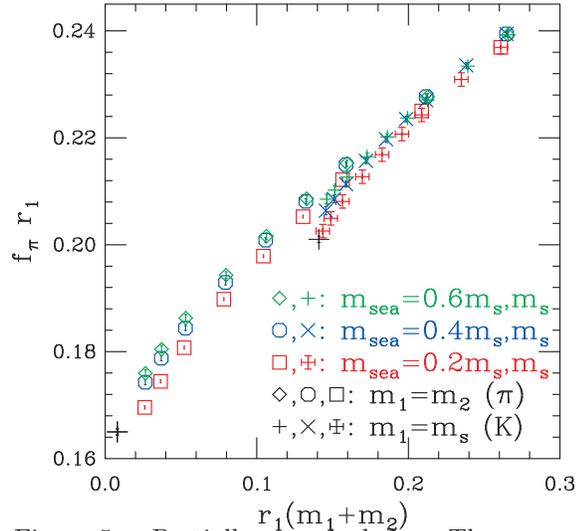} 
\rule{0.0in}{0.01in}\vspace{-0.7in}\\
\caption{
\label{FPI_R1_FIG}
Partially quenched $f_{PS}$.
These were obtained from point-point propagators, with $a \approx 0.13$ fm.
The ``+'' marks the physical $f_\pi$ and $f_K$, using $r_1=0.35$ fm.
}
\end{figure}




The MILC collaboration is engaged in a program of QCD simulations
using three flavors of dynamical quarks --- two light and one strange.
These simulations use the ``$a^2_{tad}$'' action for staggered quarks,
which removes the tree level order $a^2$ discretization errors\cite{IMP_ACTION}.
Results for scaling of hadron masses, for dynamical quark effects on
the static quark potential, and for the light hadron spectrum at
$a \approx 0.13$ fm have been reported in Refs.~\cite{ASQPUBS}.
We are continuing these simulations at smaller quark masses and on finer
lattices, with $a \approx 0.09$ fm.  These new runs allow us to extend
our studies of the approach to the continuum limit, and to study some
new problems such as exotic hybrid mesons with staggered quarks.
Several projects based on lattices from these simulations were reported
in other talks at this conference\cite{OTHERTALKS}.

Before exploring more technical points, we would like to emphasize that
the big picture of hadronic physics coming from lattice simulations is
encouraging.  For example, in Fig.~\ref{BIGPIC_FIG} we show some hadron
masses from our three flavor $a \approx 0.13$ fm calculations compared to
experimental results.  In this plot the masses (squared masses for the
pseudoscalars) were simply extrapolated linearly to the physical light
quark mass, and no continuum extrapolation was made.  Here the $\rho$
mass was used to set the scale, and the $\pi$ mass to set the light
quark mass.
(We think that the $a_1$ and $b_1$ energies are small because of couplings
to two meson states, and we are investigating this further.)

A preliminary indication of the effect of nonzero lattice spacing (the continuum
extrapolation) is seen in Fig.~\ref{MRHONUC_FIG}, where the nucleon
and vector meson ($\rho$) masses are plotted in units of $r_1$.
The leading corrections are expected to be proportional to $a^2 g^2$, or about half
the size in the 0.09 fm runs as in the 0.13 fm runs.  One can see that
the corrections are small, but possibly larger in three flavor QCD than
in the quenched approximation.

We have used the static quark potential to fix the lattice spacing in our
simulations.  The potential is also useful in phenomenological models
of hadrons.  In our earlier simulations we showed clear effects of sea
quarks on the shape of this potential.  With our 0.09 fm runs in progress,
we can begin to investigate the nonzero lattice spacing corrections to
this shape.  Figure~\ref{POTSHAPE_FIG} shows one shape parameter,
$r_0/r_1$.  ($r_0$ and $r_1$ are defined by $r^2 F(r)=1.65$ and $1.00$
respectively.)  The horizontal axis is basically the dynamical quark mass, with
the quenched limit at the right and the physical point at 0.033.  The kink at
0.45 occurs because for quark masses larger than this we varied all three
quark masses together, while for light quark masses smaller than this the
strange quark mass was held fixed.  The crosses are preliminary 0.09
fm results.  We can see that there is a small upward shift as the lattice
spacing is lowered, leading to a very crude estimate of a shift of 0.02
between $a=0.13$ fm and $a=0$.

The relatively small quark masses accessible in simulations with staggered
quarks allow us to clearly see the effects of chiral logarithms.  As Sharpe and
Shoresh have emphasized\cite{SHARPESHORESH}, partially quenched calculations
can be used to determine parameters in the chiral lagrangian that describes the
interactions of low energy pions.  Figures~\ref{MPISQ_PQ_FIG} and \ref{FPI_R1_FIG}
show partially quenched pseudoscalar masses and decay constants as a function of
valence quark mass for several values of the sea quark mass.  Curvature from
chiral logarithms is clearly visible, as is the effect of varying the sea
quark mass.  Fitting this data will require taking into account the effects
of the remaining flavor symmetry breaking, as discussed by Claude Bernard
in these proceedings\cite{BERNARD}.



\section*{Acknowledgments}
We thank Peter LePage, Junko Shigemitsu and Matt Wingate for straightening
out the normalization of $f_\pi$.
Computations were performed at
LANL, NERSC, NCSA, ORNL, PSC and SDSC.
This work was supported by the U.S. DOE and NSF.

\end{document}